\documentclass[12pt]{article}
\usepackage{cite}
\usepackage{amsmath}
\usepackage{authblk}

\usepackage[utf8]{inputenc}

\usepackage{color}

\title{A short note on gravity with tensor auxiliary fields.}
\author[1]{M\'{a}ximo Ba\~{n}ados\thanks{maxbanados@fis.puc.cl}}
\author[2]{Diego Cohen\thanks{cohen.diego@gmail.com}}
 \affil[1]{Instituto de F\'{\i}sica, P. Universidad Cat\'{o}lica de Chile, Casilla 306, Santiago 22,Chile.}
\affil[2]{Facultad de Ciencias, Universidad de Chile, Las Palmeras 3425, \~Nu\~noa, Santiago, Chile.}
 
\begin{document} 
 
 \maketitle

\begin{abstract}
We consider gravity coupled to a second metric in the strong coupling limit, where the second kinetic term is absent. This system belongs to the recently discussed class of models of ``gravity with auxiliary fields" by Pani et al.  We prove that, in vacuum, these theories are always equivalent to GR with a cosmological constant, even in the case where the auxiliary field equations contain identities leaving undetermined functions. In the situation where some functions are undetermined, the actual value of the cosmological constant is dictated by an initial condition, and not by the parameters in the action.     
\end{abstract}
 
Since the discovery of the accelerated expansion of the universe, 
huge efforts have been dedicated to study modifications and extensions of General Relativity. See \cite{Clifton:2011jh} for a recent and complete review.  In this context massive gravity, first studied by Fierz and Pauli \cite{Fierz:1939ix}, has become again a popular topic starting  with the work \cite{deRham:2010ik}.

The original theory developed in \cite{deRham:2010ik} considered Eintein's theory coupled to a second backgroud metric thus breaking diffeomorphisms invariance. It was soon realized, however, that the formulation of \cite{deRham:2010ik} could be extended to a full bigravity theory carrying 7 = 2+5 degrees of freedom, describing a massless graviton and a massive graviton with no Boulware-Deser mode.    

The action is
\begin{equation}\label{action0}
I_{\kappa} [g_{\mu\nu},f_{\alpha\beta}]  = \int_M \left(  \sqrt{g} R(g) + {1 \over \kappa}\sqrt{f}R(f) - U(g,f)  \right),
\end{equation} 
where $U(g_{\mu\nu} ,f_{\alpha\beta}) $ is a scalar density interaction depending on both fields (This action was indeed written a long time ago in \cite{ISS}). We have normalized Newton's contant to $16\pi G=1$, and leave $\kappa$ as an independent parameter. The main point raised in \cite{deRham:2010ik} is that the Boulware-Deser mode is eliminated by a particular choice of potential $U(g,f)$. See \cite{Martin-Moruno:2013wea} for a recent discussion on the relation between massive gravity of \cite{deRham:2010ik} and bigravity.

We shall be interested here in the limit $\kappa\rightarrow\infty$ obtaining the strong coupling limit of (\ref{action0}),
\begin{equation}\label{action1}
I_{\infty} [g_{\mu\nu},f_{\alpha\beta}]  = \int_M( \sqrt{g} R(g) - U(g,f)),
\end{equation}
but keeping $f_{\mu\nu}$ as a fluctuating field to maintain general covariance. For generality we shall also keep $U(g,f)$ arbitrary. Our main results will not depend on the form of the interaction piece.  

The theory described by the action (\ref{action1}) belongs to the recently discussed models of ``gravity with auxiliary fields" \cite{Pani}. The equations of motion for $f_{\mu\nu}$ are 
\begin{equation}\label{dU} 
{\partial U(g,f) \over\partial  f_{\alpha\beta }}=0,
\end{equation} 
and since $U(g,f)$ has no derivatives in $f_{\mu\nu}$, this is a set of algebraic equations for $f_{\mu\nu}$. If equations (\ref{dU}) can be solved to express $f_{\mu\nu}$ as functions of $g_{\mu\nu}$, 
\begin{equation}\label{qg}
f_{\alpha\beta} = f_{\alpha\beta}(g_{\mu\nu} ),   
\end{equation}
then plugging (\ref{qg}) back into (\ref{action1}) one obtains
\begin{equation} 
I[g_{\mu\nu}]  = \int_M( \sqrt{g} R(g) - U'(g_{\mu\nu})), 
\end{equation}     
where $U'(g) = U(g,f(g))$. Since the only scalar density build from $g_{\mu\nu} $ is $\sqrt{g}$ one concludes
\begin{equation}\label{Lambda} 
U'(g) = \Lambda \sqrt{g}, 
\end{equation}   
where $\Lambda$ is some constant depending on the parameters present in $U$.  Thus, under these conditions, the action (\ref{action1}) is fully equivalent to general relativity with a cosmological constant, for any choice of potential $U(g,f)$ \cite{Pani}.  The actual value of $\Lambda$ depends on the structure of $U$. We shall now see that theory is richer and contain branches where $\Lambda$ arises as an integration constant.

The goal of this short note is to fill up a gap in the above argument. The above reasoning is true provided the equations (\ref{dU}) can be solved, and that the expressions (\ref{qg}) actually exist. Of course we are not concerned about existence of explicit solutions. What we mean is that equations (\ref{dU}) may contain relations among them. In that case there are less equations and $f_{\mu\nu} $ cannot be solved from (\ref{dU}).   Some of its components will remain as arbitrary functions.  
In this situation, the above analysis needs to be refined. 

We shall prove two things. First, we shall prove that, even in the situation where (\ref{dU}) contain identities and the full  $f_{\mu\nu}$  cannot be expressed as a function of $g_{\mu\nu}$, this  theory is nevertheless equivalent to general relativity with a cosmological constant $\Lambda$. In this situation, however, $\Lambda$ will appear as an integration constant fixed by initial conditions, and not by the parameters in the action. Second, we shall prove that even for simple potentials, equations (\ref{dU}) do contain identities. One can expect it to be a generic property and not an isolated situation. 

Our starting point is the classification of potentials  discussed in \cite{DamourKogan}.  As argued in that reference, all scalars densities $U(g,f)$ can be written as
\begin{equation}\label{V1}
U(g,f) = \sqrt{g}\, V(g,f),
\end{equation} 
where (in 4 dimensions) $V(g,f)$ is a scalar function of the 4 scalars, 
\begin{eqnarray}
\lambda_1 &=&  g^{\mu_1\mu_2}f_{\mu_2\mu_1},  \\
\lambda_2 &=& g^{\mu_1\mu_2}f_{\mu_2\mu_3}g^{\mu_3\mu_4}f_{\mu_4\mu_1},      \\
\lambda_3 &=& g^{\mu_1\mu_2}f_{\mu_2\mu_3}g^{\mu_3\mu_4}f_{\mu_4\mu_5}g^{\mu_5\mu_6}f_{\mu_6\mu_1},   \\
\lambda_4 &=& g^{\mu_1\mu_2}f_{\mu_2\mu_3}g^{\mu_3\mu_4}f_{\mu_4\mu_5}g^{\mu_5\mu_6}f_{\mu_6\mu_7}g^{\mu_7\mu_8}f_{\mu_8\mu_1}.    
\end{eqnarray}
Higher order traces are related to these by the Caley-Hamilton theorem. Often $\sqrt{g}$ is replaced by $g^{{1 \over 4}}f^{{1 \over 4}}$, maintaining a symmetry between both fields. For our purposes it is more convenient to pull out $\sqrt{g}$. This can always be done multiplying $V(g,f)$ by a certain function of the scalars above. The potential (\ref{V1}) does represent the general situation. 

The equations of motion following from the action (\ref{action1}) with the potential (\ref{V1}) are
\begin{eqnarray}
G_{\mu\nu} &=&  {\partial V \over\partial g^{\mu\nu} } - {1 \over 2}\, V\, g_{\mu\nu}  \label{dgV},  \\
 {\partial V \over\partial f_{\mu\nu} }  &=& 0  \label{dqV}. 
\end{eqnarray} 

Let us start by proving that, for all solutions of (\ref{dqV}), the right hand side of (\ref{dgV}) is  proportional to $g_{\mu\nu}$, and thus a cosmological constant. The crucial point here is that we do not assume that (\ref{dqV}) can be solved for $f_{\mu\nu}$.   

The proof is based on the following simple identity\footnote{Exact solutions to bigravity theories have been known for a long time \cite{IshamStorey},\cite{Berezhiani:2008nr}, and they exhibit curious properties, as discussed in \cite{Banados:2011hk}. The identity (\ref{eq}), providing a striking relation between the contributions to the energy momentum tensor from both metrics, may shed some light on this issue. We plan to discuss this point elsewhere. }, 
\begin{equation}\label{eq}
g^{\mu\nu}{\partial V\over \partial g^{\nu\rho} } =  {\partial V \over \partial f_{\mu\nu}} f_{\nu\rho},
\end{equation} 
valid for any function of the invariants $\lambda_i$. Let us explore the consequences of (\ref{eq}), then we proceed with the proof. 

Since the equation of motion for $f_{\mu\nu}$ imply $ {\partial V \over \partial f_{\mu\nu}}=0$, the equality (\ref{eq}) implies that, on any solution,  ${\partial V\over \partial g^{\nu\rho} }=0$. Replacing in (\ref{dgV}), the Einstein equation is reduced to 
\begin{equation}\label{ein}
G_{\mu\nu} = - {1 \over 2}V g_{\mu\nu}.  
\end{equation}  
Since both $g_{\mu\nu}$ and $G_{\mu\nu}$ are covariantly conserved, we conclude that the coefficient of $g_{\mu\nu} $ in the right hand side must be a constant, 
\begin{equation}\label{L}
 {1 \over 2}V \equiv \Lambda,
\end{equation} 
and we recover Einstein gravity with a cosmological constant, as promised. It is important to realize that the potential $V$ evaluated on a solution to (\ref{dqV}) is not necessarily a constant. Even more, $V$ can depend on arbitrary undetermined components of $f_{\mu\nu}$ (see below). However, Bianchi identities of $G_{\mu\nu}$    put restrictions on those undetermined functions such that $\partial_\mu V$=0. In this sense the cosmological constant appear as an integration constant\footnote{See also \cite{Henneaux:1989zc}.}, and not as a fundamental parameter. 

We emphasize again that in deriving (\ref{ein}) we did not assume at any point that the equation for the auxiliary field (\ref{dqV}) imply a relation of the form (\ref{qg}). This is the main difference of our analysis as compared with \cite{Pani}. 

Let us now prove (\ref{eq}). To avoid long expressions, we consider the case where $V$ is a function of $\lambda_1$ and $\lambda_2$. The general case, depending on all 4 scalars, proceeds in exactly the same way. If $V$ dependes only on $\lambda_1$ and $\lambda_2$ applying the chain rule one obtains 
\begin{equation}\label{e1} 
{\partial V \over \partial f_{\mu\nu}}={\partial V \over \partial \lambda_1 }\,\, g^{\mu\nu} + 2{\partial V \over \partial\lambda_2 }\,\, g^{\mu\alpha}f_{\alpha\beta}g^{\beta\nu}.
\end{equation}    
In the same way, 
\begin{equation}\label{e2} 
{\partial V \over \partial g^{\mu\nu} } ={\partial V \over\partial \lambda_1 }\, f_{\mu\nu} + 2{\partial V \over\partial \lambda_2 }\, f_{\mu\alpha}g^{\alpha\beta}f_{\beta\nu}.    
\end{equation} 
Multiplying (\ref{e1}) by$f_{\nu\sigma} $ and (\ref{e2}) by $g^{\mu\rho}$ one readily obtains (\ref{eq}).

Let is finally discuss the most important point of this note, namely, that generically the equations (\ref{dqV}) for the auxiliary field may contain identities and hence a relation of the form (\ref{qg}) may not exist at all.  

Consider, as an example, the simple interaction potential 
\begin{equation}\label{V2} 
U(g,f)  = \sqrt{g}\, (g^{\mu\nu}f_{\mu\nu} + \alpha \, f^{\mu\nu}g_{\mu\nu}). 
\end{equation} 
Here, $f^{\alpha\beta}$ is the inverse of $f_{\beta\rho}$ and $\alpha$ is a constant.  We shall see that this simple example exhibits branches where equations (\ref{dqV}) have identities. One can then conjecture that this is a generic property of tensor potentials and not an isolated case.  

The equations of motion for the field $f_{\mu\nu}$ are 
\begin{equation}
{1 \over \sqrt{g}}{\partial U \over \partial f_{\mu\nu}}= g^{\mu\nu} - \alpha\, f^{\mu\alpha}g_{\alpha\beta}f^{\beta\nu}   =0.
\end{equation} 
Multiplying by $f_{\mu\rho} f_{\nu\sigma}$ these equations can be written in a convenient way as  
\begin{equation}\label{eqq} 
g^{\mu\nu} f_{\mu\rho}f_{\nu\sigma} = \alpha\, g_{\rho \sigma}.
\end{equation}  
Our problem is to solve these equations for $f_{\mu\nu}$.  

One solution to (\ref{eqq}) is simply $f_{\mu\nu} = \alpha^{1/2}g_{\mu\nu}$ which does determine $f_{\mu\nu}$ completely in terms of $g_{\mu\nu}$. Plugging back this solution into (\ref{V2}) we immediately see that $U$ becomes a cosmological constant term. It is useful to note that, in general, for proportional solutions, $f_{\mu\nu}=x\,g_{\mu\nu}$, the values of all invariants $\lambda_i$ are real numbers. One can clearly see that for this class of solutions, the value of $\Lambda$ will be a combination of the parameters originally present in the potential.   

However, the interesting point is that the proportional field is  {\it not} the most general solution. Other branches exist where some functions are undetermined and the value of $\Lambda$ appears as an initial condition.  

To keep the discussion simple let us look at equation (\ref{eqq}) in 2 dimensions. Since we would like to find $f_{\mu\nu} $ for a given $g_{\mu\nu} $, let us choose the simplest diagonal $g_{\mu\nu}$, 
\begin{equation} 
g_{\mu\nu}  = \left(  \begin{array}{cc}
g & 0 \\ 
0 & g
\end{array}  \right).
\end{equation}     
On the other hand we write a general ansatz for a symmetric $f_{\mu\nu} $
\begin{equation}
f_{\mu\nu} = \left( \begin{array}{cc}
x & y \\ 
y & z
\end{array}  \right).
\end{equation}    
We plug these fields into (\ref{eqq}) and obtain 3 equations,
\begin{eqnarray}
x^2+y^2-\alpha g^2 &=& 0 \label{1}, \\  
z^2+y^2-\alpha g^2 &=& 0 \label{2}, \\  
y(x+z) &=& 0  \label{3}.  
\end{eqnarray} 
Note that equation (\ref{3}) has two branches. If one chooses $y=0$, then equations (\ref{1}),(\ref{2}) fix $x,z$ as functions of $f$, as expected.  This in fact is equivalent to the proportional solution described above. However, if one chooses the second branch,    
\begin{equation}
x+z=0,
\end{equation} 
we observe that equations (\ref{1}) and (\ref{2}) become the same. Therefore, there remains only one equation for two variables. On this second branch, there are identities between equations (\ref{1}),(\ref{2}) and (\ref{3}), and they do  not fix the tensor $f_{\mu\nu}$ completely.  

As showed above, the dynamics of the metric $g_{\mu\nu}$ is nevertheless governed by Einstein equations with a cosmological constant. But the actual value of $\Lambda$ is not determined by a parameter in the action, but as an initial condition.  Furthermore, the properties of the field $f_{\mu\nu}$ are very different in each branch. On the proportional branch $f_{\mu\nu}$ becomes completely fixed in terms of $g_{\mu\nu}$. On the other non-proportional branch ($y\neq 0$) some of its components remain arbitrary. This is a signal of a ``hidden" gauge symmetry, which is we shall explore elsewhere.  

To conclude, in this short note we have proved that gravity coupled to a second metric $f_{\mu\nu}$, in the strong regime where the kinetic term for it is absent, is always equivalent to general relativity with a cosmological constant. Even in the case where the equations for $f_{\mu\nu}$ contain identities and do not fix that field completely. Adding a matter Lagrangian $L_{m}(g_{\mu\nu})$ clearly does not change the conclusions of this note. 

~ 

The authors would like to thank Paolo Pani for a detailed reading of the manuscript and several useful comments. 
The authors are partially supported by Fondecyt (Chile) \#1100282 and Anillo ACT (Chile)  \#1102.


 \end{document}